\documentclass[a4paper, 12pt]{article}

\usepackage[ansinew]{inputenc}
\usepackage{amsmath}
\usepackage{graphicx}

\begin{document}

\newcommand{\sign}{\operatorname{sign}}
\newcommand{\Ci}{\operatorname{Ci}}
\newcommand{\Si}{\operatorname{Si}}
\newcommand{\tr}{\operatorname{tr}}

\newcommand{\beq}{\begin{equation}}
\newcommand{\eeq}{\end{equation}}
\newcommand{\beqn}{\begin{eqnarray}}
\newcommand{\eeqn}{\end{eqnarray}}

\newcommand{\slp}{\raise.15ex\hbox{$/$}\kern-.57em\hbox{$ \partial $}}
\newcommand{\lnA}{\raise.15ex\hbox{$/$}\kern-.57em\hbox{$A$}}
\newcommand{\unmedio}{{\scriptstyle\frac{1}{2}}}
\newcommand{\uncuarto}{{\scriptstyle\frac{1}{4}}}

\newcommand{\trial}{_{\text{trial}}}
\newcommand{\true}{_{\text{true}}}
\newcommand{\const}{\text{const}}

\newcommand{\intp}{\int\frac{d^2p}{(2\pi)^2}\,}
\newcommand{\intx}{\int_C d^2x\,}
\newcommand{\inty}{\int_C d^2y\,}
\newcommand{\intxy}{\int_C d^2x\,d^2y\,}

\newcommand{\bP}{\bar{\Psi}}
\newcommand{\bc}{\bar{\chi}}
\newcommand{\hs}{\hspace*{0.6cm}}

\newcommand{\bra}{\left\langle}
\newcommand{\ket}{\right\rangle}
\newcommand{\bracket}{\left\langle\,\right\rangle}

\newcommand{\D}{\mbox{$\mathcal{D}$}}
\newcommand{\N}{\mbox{$\mathcal{N}$}}
\newcommand{\Lag}{\mbox{$\mathcal{L}$}}
\newcommand{\V}{\mbox{$\mathcal{V}$}}
\newcommand{\Z}{\mbox{$\mathcal{Z}$}}
\newcommand{\A}{\mbox{$\mathcal{A}$}}
\newcommand{\B}{\mbox{$\mathcal{B}$}}
\newcommand{\C}{\mbox{$\mathcal{C}$}}
\newcommand{\E}{\mbox{$\mathcal{E}$}}


\vspace{2cm}

\begin{center}

{\Large {\bf Effect of non adiabatic switching of dynamic
perturbations in 1d Fermi systems}}

\vspace{1.3cm}

Carlos M. Naón$^{a,b}$, Mariano J. Salvay$^{a,b}$, Marta L.
Trobo$^{a,b}$ \footnote{e-mail: salvay@fisica.unlp.edu.ar,
naon@fisica.unlp.edu.ar, trobo@fisica.unlp.edu.ar}

\vspace{.8cm}

$^a$ {\it Instituto de Física La Plata, Departamento de Física,
Facultad de Ciencias Exactas, Universidad Nacional de La Plata.
CC 67, 1900 La Plata, Argentina.}

\smallskip

$^b$ {\it Consejo Nacional de Investigaciones Científicas y
Técnicas, Argentina.}

\vspace{1cm}

\begin{abstract}
We study a two-dimensional fermionic QFT used to model 1D strongly
correlated electrons in the presence of a time-dependent impurity
that drives the system out of equilibrium. In contrast to previous
investigations, we consider a dynamic barrier switched on at a
finite time. We compute the total energy density (TED) of the
system and establish two well defined regimes in terms of the
relationship between the frequency of the time-dependent
perturbation $\Omega$ and the electron energy $\omega$. Finally,
we derive a relaxation time $t_{R}$ such that for times shorter
than $t_{R}$ the finite-time switching process is relevant.
\end{abstract}
\end{center}

\vspace{1 cm}


\noindent{\it Pacs:} 71.10.Pm, 05.30.Fk

\newpage

Low dimensional field theories are objects of sustained interest.
In particular they are useful to describe the behavior of strongly
anisotropic physical systems such as organic conductors
\cite{organic conductors}, charge transfer salts \cite{salts},
quantum wires \cite{quantum wires}, and carbon nanotubes
\cite{CNT}. In this condensed matter context, one of the most
widely studied 1D theory is the so called "g-ology" model
\cite{Solyom}, which displays the Luttinger liquid behavior
characterized by spin-charge separation and interaction-dependent
power-law correlation functions \cite{Haldane}. Many of the
advances in the understanding of this model have been accomplished
through the use of the renormalization group and the bosonization
procedure, and in the context of equilibrium situations, i.e. in
the absence of time-dependent interactions. However, as it is well
known, in nature, thermal equilibrium is more an exception than a
rule. In fact, in recent years there has been an increasing
activity focused on the study of out of equilibrium
low-dimensional electronic systems \cite{Chamon}. Thus, it becomes
natural to consider the formulation of field-theoretical methods
that can be applied to this kind of systems. To this aim, we have
recently shown how to combine the Closed Time Path (CTP) method of
Schwinger and Keldysh \cite{CTP} with a path-integral approach to
bosonization \cite{Nosotros} in order to analyze the g-ology model
in the presence of a time-dependent dynamic barrier \cite{NST}. On
the other hand, as far as we know, all previous works on 1D
electronic systems with time-dependent impurities (including
ref.\cite{NST}) use a large-time approximation in order to treat
the correlation functions. In this case the effect of a non
adiabatic switching of the interaction can be disregarded and the
total energy density (TED) becomes an harmonic function of time
which temporal average can be easily defined and computed. An
obvious and drastic consequence is that no transient process can
be predicted. The main purpose of this work is to reconsider this
problem by taking into account the role of a sudden, finite-time
switching process. This will enable us to show how the TED evolves
from short to long-time regimes.

We are interested in computing the TED for a Luttinger system
defined by

\begin{equation} S = S_{0} + S_{int} + S_{I},\label{a}
\end{equation}
where $S_{0}$ is the unperturbed action (in the condensed matter
context it is thought of as a linearized free dispersion
relation):
\begin{equation}
 S_{0} = \intx\bP i\slp\Psi ,
\end{equation}
where $\int_C$ indicates that the integration is defined along the
time contour usually defined in the CTP formalism \cite{CTP}.
$S_{int}$ describes the forward scattering of spinless fermions
(electrons):
\begin{equation}
 S_{int} = -\frac{1}{2} \intxy ( \bP \gamma_{\mu} \Psi )
(\textbf{x})~ V_{(\mu)}(\textbf{x}-\textbf{y}) ~( \bP \gamma_{\mu}
\Psi ) (\textbf{y})
\end{equation}
where $\textbf{x} = (x, t)$ and the fermionic currents $\bP
\gamma_{\mu} \Psi$  are coupled through distance-dependent
functions, $V_{(\mu)}(\textbf{x}-\textbf{y})$. In terms of these
potentials one can make direct contact with the forward-scattering
sector of the "g-ology" model currently used to describe different
scattering processes characterized by coupling functions $g_1$,
$g_2$, $g_3$ and $g_4$ \cite{Solyom}. Neglecting processes
associated to large momentum transfers, only the forward
scattering couplings $g_2$ and $g_4$ play a role. The relation
between these strengths and our potentials are given (in Fourier
space,$\textbf{p} = (p, \omega)$ ) by
\begin{eqnarray}
g_{2}(\textbf{p}) & = \frac{1}{2}(V_{(1)}(\textbf{p}) +
V_{(0)}(\textbf{p})) \nonumber
\\ g_{4}(\textbf{p}) & = \frac{1}{2}(V_{(0)}(\textbf{p}) -
V_{(1)}(\textbf{p})).
\end{eqnarray}
From now on we shall disregard the momentum dependence of the
potentials, i.e. only short-range interaction will be taken into
account. The action $S_{I}$ describes the interaction between the
electrons and a localized time-dependent perturbation:
\begin{equation}
S_{I} = \intx\bP \,\gamma_{0} \,\Delta (x, t )\,\Psi ,
\end{equation}
where $\gamma_{0}$ is a Dirac matrix and $\Delta (x, t )$ is a
function that contains the details of the perturbation, i.e. the
way in which the interaction is switched on in time and the form
in which it is localized in space. Please note that, for
simplicity, only forward scattering between the electron and the
impurity is considered. As we shall see later, this is not a bad
approximation if weak couplings and short times are taken into
account. Besides, in some experimental arrangements one can
reasonably argue that backscattering effects will not be important
(see, for instance \cite{Poncharal}). To be specific we choose a
particular form for the perturbation:

\begin{equation}\label{perturbation}
\Delta(x, t ) = \lambda \, \sin(\Omega t )\, \Theta( t )[ \Theta(
x + a/2 ) - \Theta( x - a/2 )]
\end{equation}
where $\Theta(t)$ is the Heaviside function. We then have a
separable harmonic perturbation suddenly switched on, consisting
of a barrier of width $a$ and height $\lambda$ that oscillates in
time with frequency $\Omega$. As we shall see, this choice allows
us to make contact with previous works on the effect of
time-dependent perturbations on one-dimensional systems. Indeed, a
similar potential (with an additional static term) was considered
by B\"{u}ttiker and Landauer in their study of the traversal time
for tunneling \cite{BL}. More recently, and in the context of
Luttinger liquids, temporal harmonic perturbations were analyzed
for both forward \cite{Komnik} and backscattering \cite{Gefen}
impurities. Let us stress that neither of these authors includes
the effect of switching, which is our main motivation.

Let us now focus our attention on the TED, which in the Wigner
representation can be expressed in terms of the correlation
function as
\begin{equation}
n(\omega, x, t) = - i \int_{-\infty}^{\infty} d \tau \exp ( i
\omega \tau )\,G_{- +}(x, x, t + \tau/2 , t - \tau/2),\label{C}
\end{equation}
where the right and left fermionic propagators are time-ordered
along a time contour C:
\begin{eqnarray}\label{Greens}
G^{R,L}_C =  \left(\begin{array}{cl} G^{R,L}_{++} & G^{R,L}_{+-}
\\ \\ G^{R,L}_{-+} & G^{R,L}_{--} \end{array} \right).
\end{eqnarray}
The subscripts $+$ and $-$ refer to fields defined in the upper
and lower branches of C, respectively, corresponding to forward
$(+)$ and backward $(-)$ time evolution. From now on we shall
restrict our analysis to the TED corresponding to right-movers.
Similar results can be obtained for left-moving particles.

Using the technique described in \cite{NST} we can factorize the
Green function as
\begin{equation}\label{b}
G_{C}(x, y, t, t^{'}) = G_{C}^{\gamma}(x - y, t - t^{'}) \exp
\bigl[ \beta (x, t) - \beta (y, t^{'})\bigl],
\end{equation}
where the function $G_{C}^{\gamma}(x - y, t - t^{'})$ is the
equilibrium propagator for a Luttinger liquid with its
characteristic interaction dependent exponent $\gamma=(1/4) ( K +
K^{-1} - 2)$, ($K  =  \sqrt{\frac{1+ g_{4}/ \pi -g_{2}/ \pi}{1+
g_{4}/ \pi + g_{2}/ \pi}}$ ) and
\begin{equation}
\beta(x,t) = \frac{-i \lambda}{2\Omega} \Theta(t) \{ e^{i \Omega t
} F(x,a,\Omega) + e^{- i \Omega t } F(x,a,- \Omega)\}
\end{equation}
with
\begin{eqnarray} F( x , a ,\Omega)& = \Theta ( - ( x + a/2) )\exp
[ - i \Omega ( x + a/2 ) ] - \Theta ( - ( x - a/2) )\exp [ - i
\Omega ( x - a/2 ) ] + \nonumber \\& +  \Theta ( x + a/2 ) -
\Theta (  x - a/2 ).
\end{eqnarray}

Since we are mainly concerned with the effect of the switching
process, we start by studying the free case in the presence of the
time-dependent impurity ($\gamma =0$). For this case the
correlation function is given by
\begin{equation}
G^{(0)}_{-+}(x , t + \tau/2; x, t - \tau/2) = \frac{i}{2 \pi}
\frac{1}{\alpha_0 +  i\tau} \label{B}.
\end{equation}
Expanding the exponential in eq. (\ref{b}), expressing the
$\beta$-dependent factor in terms of the binomial expansion and
replacing (\ref{b}) in (\ref{C}), after some algebra one can write
the TED as
\begin{multline}\label{libre}
n(\omega,x,t)_{\gamma=0} = \Theta(\omega) + \sum_{n=1}^{\infty}
\Bigl( \frac{-i \lambda}{2 \Omega}\Bigr)^n \sum_{j=0}^{n}
\frac{1}{ (n-j)! j!}\bigl( F(x,a,\Omega)\bigr)^{n-j} \bigl(F(x,a,-
\Omega)\bigr)^j \exp\{ i \Omega t(n-2j)\}\times \\
\times \Bigl[ \Theta(\omega +(n-2j)\Omega/2) + \frac{i}{2 \pi}
\bigl( \Ci [(2t + i \alpha_0)(\omega + (n-2j)\Omega/2)] - i \Si
[(2t + i \alpha_0)(\omega + (n-2j)\Omega/2)] + \\+ (-1)^{n+1}
\bigl( \Ci [(2t - i \alpha_0)(- \omega + (n-2j)\Omega/2)] - i \Si
[(2t - i \alpha_0)(- \omega + (n-2j)\Omega/2)]\bigr)\bigr) \Bigr]
+
\\+ \sum_{n=2}^{\infty} \Bigl( \frac{-i \lambda}{2 \Omega}\Bigr)^n
\sum_{j=1}^{n-1}(-1)^j \sum_{k=0}^{j} \sum_{l=0}^{n-j}
\frac{1}{(j-k)! i! (n-j-l)! l!} \bigl(F(x,a,\Omega)\bigr)^{n-l-k}
\bigl(F(x,a,- \Omega)\bigr)^{l+k} \times \\ \times \frac{i
\Theta(t)}{2 \pi} \exp\{i \Omega t (n - 2k -2l)\} \Bigl[ \Ci
\bigl[(2t + i \alpha_0)(\omega + (n-2j-2l+2k)\Omega /2)\bigr]+ \\
- i \Si \bigl[(2t + i \alpha_0)(\omega + (n-2j-2l+2k)\Omega
/2)\bigr]  - \Ci \bigl[(-2t + i \alpha_0)(\omega +
(n-2j-2l+2k)\Omega /2)\bigr] + \\ + i \Si \bigl[(-2t + i \alpha_0
)(\omega + (n-2j-2l+2k)\Omega /2)\bigr] \Bigr].
\end{multline}
Let us mention that the first term in the above formula is the
equilibrium TED. Please observe that for $t\rightarrow-\infty$,
i.e. for long times in the past, only this first term remains and
the TED recovers its equilibrium value: $n(\omega,x,t)_{\gamma=0}
= n_{0}(\omega)=\Theta(\omega)$, as expected. We stress that the
inclusion of the factor $\Theta(t)$ in the perturbation gives rise
to the appearance of non harmonic contributions to the TED. These
contributions manifest through the functions cosine-integral (Ci)
and sine-integral (Si). For large positive times only harmonic
terms survive. In this regime the TED is a harmonic superposition
of equilibrium TED's centered in integer multiples of $\Omega/2$,
with coefficients that depend on the geometry and strength of the
time-dependent interaction. Only in this regime a time average
(over the period of the interaction) does not depend on the
temporal interval and leads to an averaged TED which is a
superposition of equilibrium TED's centered in integer multiples
of $\Omega$ \cite{Komnik} \cite{NST}. Some of these features can
be globally seen in figures 1 and 2, where we show the behavior of
$n(\omega,x,t)_{\gamma=0}$ as function of $\omega$ and $t$, for
fixed $x$ inside the barrier ($-a/2 < x < a/2$). For comparison
purposes, in figure 1 we disregarded the effect of the non
adiabatic switching, whereas in figure 2 the $\Theta(t)$ function
has been included. We have set $\lambda=2$, $\Omega=1$ and $a/2
-x=1$.

\begin{figure}
\begin{center}
\includegraphics{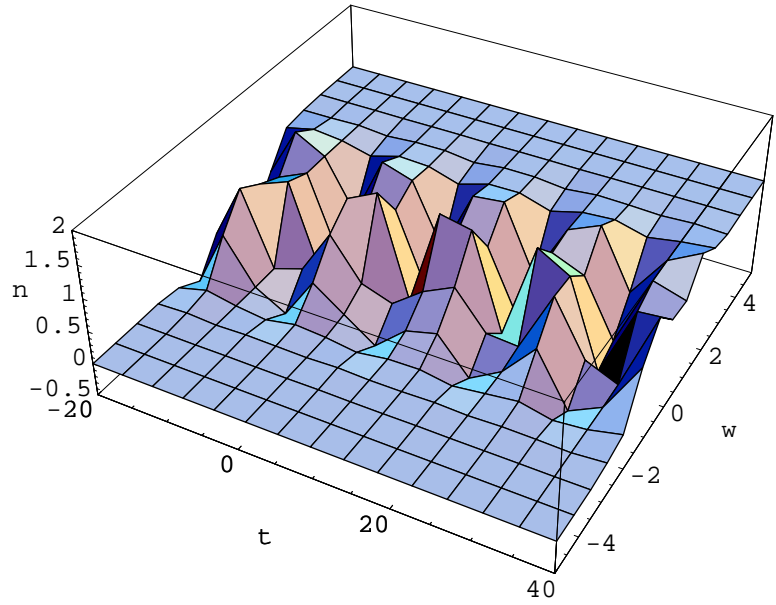}
\caption{\label{figura1}: $n(\omega,t)_{\gamma=0}$ as function of
$\omega$ and $t$, without finite-time non adiabatic switching.}
\end{center}
\end{figure}

\begin{figure}
\begin{center}
\includegraphics{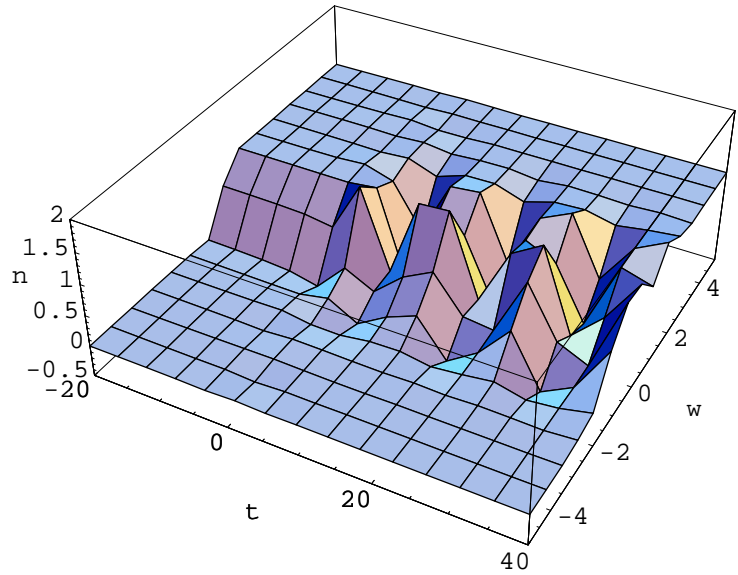}
\caption{\label{figura2}: $n(\omega,t)_{\gamma=0}$ as function of
$\omega$ and $t$, with finite-time non adiabatic switching.}
\end{center}
\end{figure}

When one focus the attention on the behavior of the TED as the
energy $\omega$ is varied, another interesting feature is
revealed. We can establish two well defined regimes in terms of
the relationship between the external frequency $\Omega$ and the
electron energy $\omega$. Indeed, for $\Omega/\omega\ll1$ and
positive times, the electrons "do not see" the harmonic
perturbation and the corresponding TED is equal to
$n_{0}(\omega)$. So, this is the large energy region where the
effect of the impurity is negligible. At this point we should
emphasize that the impurity considered in this paper does not
include a static term. Had we considered such a contribution we
would have found that the large energy regime is dominated by the
static impurity. The transient effects arising in this case have
been extensively investigated in connection to the X-ray problem
and the orthogonality catastrophe, for both Fermi
\cite{X-ray-ortho-Fermi} and Luttinger liquids
\cite{X-ray-ortho-Luttinger}. In our framework, however,in order
to make a precise description of the X-ray problem, one should
modify the temporal behavior of the impurity, i.e. the function
associated to the switching process, such that a switching off
time is also incorporated.

Going back to the analysis of our TED, one sees that for
$\Omega/\omega\gg1$ it is greatly affected by the impurity. In the
absence of switching (figure 1) the electrons gain or lose energy
quanta of value $\pm n\Omega$ (this is clearly seen when one
considers the time average). The same occurs when switching is
taken into account (figure 2), for sufficiently large times. Of
course, there is also a crossover region given by
$\Omega/\omega\cong1$. At this point we can make contact with
previous studies on the traversal time for tunneling in the
context of the quantum mechanics of a particle in the presence of
a dynamic barrier\cite{BL}. From the uncertainty principle we know
that the traversal time $\tau$ satisfies $\tau\sim1/\omega$, which
leads to $\tau \Omega \cong1$, a result which is consistent with
the findings of ref.\cite{BL}.

\begin{figure}
\begin{center}
\includegraphics{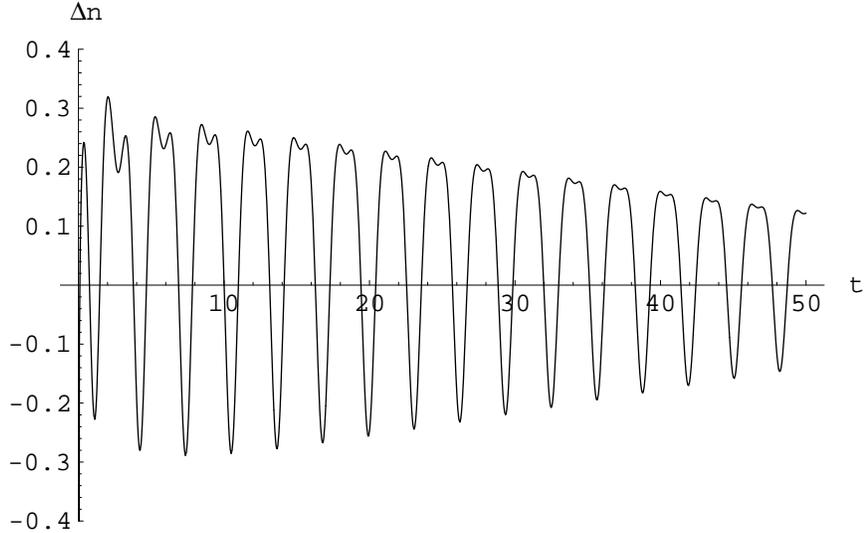}
\caption{\label{figura3}: $\Delta n(t)$ in the low energy regime,
for $\gamma=0$, $\Omega=1$, $\omega=0.01$.}
\end{center}
\end{figure}

In order to perform a quantitative description of the transient
process which is now accesible due to the introduction of a
finite-time switching mechanism, we found useful to consider the
difference between TED's with and without the inclusion of the
temporal Heaviside function. We call this difference $\Delta
n(t)$. In Figure 3 we plot $\Delta n(t)$ in the low energy regime
($\Omega=1$,$\omega=0.01$). From this figure one can see that it
is possible to define and compute a relaxation time $t_{R}$ such
that for times larger than $t_{R}$ the TED coincides with the one
obtained when the impurity is always switched on. Thus, $t<t_{R}$
defines a temporal region in which the switching process is
relevant. We have analyzed the envelope of $\Delta n(t)$ for many
different values of $\omega$, always in the low energy region
$\Omega/\omega\gg1$. In other words, we fit the envelope with an
exponential decay $b \exp{(-t/t_{R})}$. This procedure allowed us
to find $t_{R}=1/(2\omega)$. The variation of $\lambda$ and
$\Omega$ only affects the values of $b$.

Let us now say a few words on the role of backscattering from the
impurity. If we add to the Lagrangian the corresponding term, of
the form $\bP \,\Delta_{back} (x, t )\,\Psi$, of course, the
problem of determining the Green's functions is not exactly
solvable any more and one is forced to make a perturbative
calculation. For a general interaction of the form $\Delta_{back}
(x, t )= \lambda_{back}\,f(x,t)$, where the new coupling constant
$\lambda_{back}$ is assumed to be of the same order of $\lambda$,
it is easy to show that up to first order in these perturbative
parameters, only the forward scattering term affects the behavior
of the TED. This means that in the short-time regime ($t<t_R$) and
for weak coupling, only forward scattering from the impurity
contributes to the TED. In the large-time regime, the first order
contribution in $\lambda$ gives a zero average and then one is led
to a higher order calculation in which backscattering effects
would eventually contribute in a non trivial way. A quantitative
discussion of these effects are beyond the scope of the present
work, but will be the subject of future investigations.

We have also computed the electrical density current $J$ in the
presence of the time dependent interaction studied throughout this
work. Using a standard definition of $J$ \cite{Mahan} we obtained
the following remarkably simple expression:
\begin{multline}
J(x,t)=C \lambda
\,\Theta(t)\,\Bigl(\Theta(-x-\frac{a}{2})\,\sin[\Omega(t-x-\frac{a}{2})]-
\Theta(-x+\frac{a}{2})\,\sin[\Omega(t-x+\frac{a}{2})]+\\+
\Theta(x+\frac{a}{2})\,\sin[\Omega(t+x+\frac{a}{2})]-\Theta(x-\frac{a}{2})\,\sin[\Omega(t+x-\frac{a}{2})]
\Bigr),
\end{multline}
where $C$ is a renormalization constant. Let us stress that this
is an exact result (in the absence of backscattering no
perturbative expansion is required). We see that the current
originated by the impurity (note that no external voltage has been
applied) is a simple superposition of harmonic contributions and
therefore its temporal average vanishes. If one introduces a bias
V one obtains a (time-independent) term linear in V plus a
photocurrent contribution similar to the above expression but with
additional factors that depend on the right and left chemical
potentials. The temporal average of this photocurrent is still
zero. We then conclude that the dc conductance of the system is
not affected by the perturbation. Again, this result is not
changed by backscattering from the impurity, at least up to first
order in the couplings.

Turning back to the TED, we have extended the previous analysis to
the case of a Luttinger liquid. Performing the same kind of
manipulations as in the non interacting case, we found an
expression similar to the one corresponding to the $\gamma=0$ case
(eq. (\ref{libre})). Since it is another lengthy expression we
shall not write it down here. The only relevant difference is the
appearance of decaying factors of the form $\exp(-\Lambda/v
|\omega \pm n\Omega/2|)$ ($\Lambda$ is an ultraviolet cutoff and
$v$ the renormalized velocity), characteristic of the smoothing of
the energy density as consequence of the electron-electron
interaction. One can also verify that $n_{\gamma \neq0}<n_{\gamma
= 0}$. Another interesting point we want to mention concerns the
comparison between the areas under the curves of $n$ and $n_0$
(TED's with and without the time-dependent impurity) as functions
of $t$ for both the interacting and non-interacting electrons. In
the first case the presence of the impurity leads to a smaller
value for the total area, whereas a non vanishing forward
scattering ($\gamma\neq0$) produces the opposite effect. This
occurs with and without a finite-time non adiabatic switching.

In Figure 4 we show $\Delta n(t)$ for $\gamma=1/2$, $\Lambda=0.1$
and $v=1$. We kept the same values used in Figure 3 for the other
parameters. One observes that $\Delta n(t)$ takes much lower
values due to the fact that, in the presence of electron-electron
forward-scattering, the TED's are smaller as consequence of the
decaying factors mentioned above. On the other hand, comparing
figures 3 and 4, it becomes apparent that the relaxation time is
drastically diminished by the interaction. Then we conclude that
for a Luttinger liquid with sufficiently large $\gamma$ the effect
of the finite-time switching can be safely neglected.

\begin{figure}
\begin{center}
\includegraphics{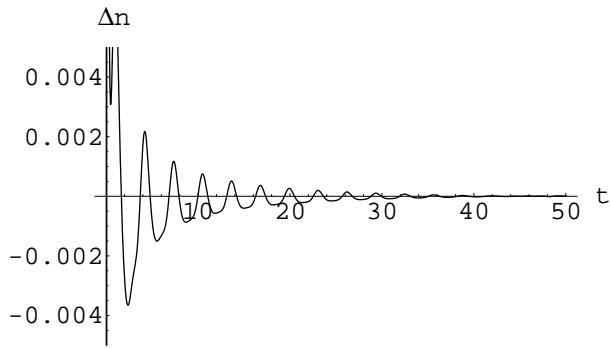}
\caption{\label{figura4}: $\Delta n(t)$ in the low energy regime,
for $\gamma=1/2$, $\Omega=1$, $\omega=0.01$.}
\end{center}
\end{figure}

\vspace{.5cm}

In summary, we have considered the effect of a harmonic
time-dependent perturbation on the TED of a 1D fermionic system.
We put special emphasis on the case in which the perturbation is
switched on suddenly at a finite time, and compared our results
with the ones obtained when the dynamic interaction is present at
all times. The low-energy transient process by which the TED
corresponding to finite-time switching evolves to the long-time
(harmonic) TED is most clearly characterized by the difference
$\Delta n(t)$. By carefully analyzing this function for different
energies $\omega$ we were able to determine that the relaxation
time $t_R$ associated to the TED evolution, for free fermions
($\gamma=0$), is equal to $1/(2\omega)$. We also established that
for Luttinger liquids ($\gamma\neq0$) $t_R$ is significantly
shortened. This suggests that, in order to observe some
consequence of the transient process, it is convenient to consider
Luttinger systems with values of $\gamma$ as small as possible.
Concerning the transport properties, we showed that the electrical
current originated by the time-dependent perturbation is a simple
superposition of harmonic contributions. This means that if one
introduces an external voltage V (through appropriate chemical
potentials) the dc conductance will remain unchanged by the
dynamic perturbation.

\section*{Acknowledgements}
This work was partially supported by Universidad Nacional de La
Plata  and Consejo Nacional de Investigaciones Científicas y
Técnicas, CONICET (Argentina).

\end{document}